\documentclass[letterpaper,titlepage,11pt]{article}

\usepackage{jheppub}

\usepackage[T1]{fontenc}
\usepackage[utf8]{inputenc} % pdfLaTeX için
\usepackage[turkish,english]{babel}

\usepackage{amsmath,amssymb,amsfonts}
\usepackage{esint}
\usepackage{empheq}
\usepackage{multirow}
\usepackage{enumerate}
\usepackage{needspace}

\usepackage{xcolor}
\usepackage{tikz}

\usepackage{textcomp}
\usepackage{bbm}

\usepackage[active]{srcltx}

\title{\boldmath\color{black}{
A Free-Field Realization of Boundary Symmetries \\ in Discrete Jackiw--Teitelboim Gravity
}}
\subheader{ITU Preprint,\,\\ \today}
\newcommand{\itu}{\dagger}

\author[\itu]{H.~T.~\"Ozer}
\emailAdd{ozert@itu.edu.tr}

\author[\itu]{Ayt\"ul~Filiz}
\emailAdd{aytulfiliz@itu.edu.tr}

\affiliation[\itu]{Istanbul Technical University,\,Faculty of Science and Letters,
\,Physics Department,\\34469 Maslak,\,Istanbul, Turkey.}
\abstract{
We investigate a finite-periodic Miura free-field realization of the classical
Virasoro--dilaton sector relevant to $\mathfrak{sl}(2,\mathbb R)$
Jackiw--Teitelboim (JT) gravity. Starting from a single logarithmic
Cartan field, we use the established discrete Miura map and its
free-field Poisson bracket to construct the lattice spin-two current.
The dilaton is then realized as a periodic composite of the same
Cartan variables, selected by an inhomogeneous difference equation and
its monodromy condition. This construction yields an explicit induced
mixed Poisson bracket without postulating an independent dilaton bracket.
We examine the associated transfer problem and quadratic adjoint
invariant, and test the finite Fourier algebra against the continuous
Virasoro--dilaton transformation, in which the dilaton has conformal
weight minus one. The construction supplies a free-field counterpart
of the JT boundary transformation laws; it does not by itself identify
the Miura Poisson structure with the canonical BF boundary-charge
algebra.
}


\begin{document}
\maketitle
%%%%%%%%%%%%%%%%%%%%%%%%%%%%%%%%%%%%%%%%%%%%%%%%%%%%%%%%%%%%%%%%%%%%%%%%%%%%%%%%%%%%%%%%%%
%%%%%%%%%%%%%%%%%%%%%%%%%%%%%%%%%%%%%%%%%%%%%%%%%%%%%%%%%%%%%%%%%%%%%%%%%%%%%%%%%%%%%%%%%%
%%%%%%%%%%      INTRODUCTION   %%%%%%%%%%%%%%%%%%%%%%%%%%%%%%%%%%%%%%%%%%%%%%%%%%%%%%%%%%%
%%%%%%%%%%%%%%%%%%%%%%%%%%%%%%%%%%%%%%%%%%%%%%%%%%%%%%%%%%%%%%%%%%%%%%%%%%%%%%%%%%%%%%%%%%
%%%%%%%%%%%%%%%%%%%%%%%%%%%%%%%%%%%%%%%%%%%%%%%%%%%%%%%%%%%%%%%%%%%%%%%%%%%%%%%%%%%%%%%%%%
\section{Introduction}\label{sec:introduction}

Jackiw--Teitelboim (JT) gravity provides a gauge-theoretic setting in
which the relation between bulk constraints and boundary symmetries can
be investigated explicitly
\cite{Jackiw1984,Teitelboim1983,FukuyamaKamimura1985,GrumillerKummerVassilevich2002}.
In the $\mathfrak{sl}(2,\mathbb R)$ BF description, the connection is
flat and the adjoint dilaton is covariantly constant. Conformal boundary
conditions nevertheless admit nontrivial residual transformations, and
the dilaton selects the transformations compatible with a given boundary
configuration \cite{Grumiller:2017qao}. The resulting spin-two
structure and dilaton transformation are especially transparent in the
continuous BF analysis of ref.~\cite{Ozer:2025bpb}. The question
addressed here is whether both can be realized from one and the same
set of discrete Miura free fields.

The gauge-covariant discretization of gravity naturally uses group-valued
transport variables rather than finite differences of a continuum
connection. The use of holonomies in lattice gauge theory has a long
history \cite{Wilson1974}, and related constructions have been developed
in discrete gravity \cite{Dupuis:2017otn}. In particular,
ref.~\cite{OzerFiliz:2026DiscreteJT} established a bulk-first,
holonomy-based discrete BF description of JT gravity, including its
boundary symmetry and dilaton sectors. Our aim is different from
repeating that construction: we seek a concrete Miura free-field
realization of the \emph{mixed} Virasoro--dilaton structure. This
requires the dilaton to be represented through the same Cartan
variables that generate the discrete spin-two current, rather than
merely being transported as an independent adjoint field.

Drinfeld--Sokolov reduction relates constrained gauge connections to
conformal algebras \cite{DrinfeldSokolov1985}. The associated Miura
construction represents the reduced currents in terms of Cartan free
fields, a connection familiar from both classical reduction and the
quantum theory \cite{FeiginFrenkel1990,BouwknegtSchoutens1993}.
The lattice analogue is not new. Faddeev and Volkov studied a lattice
Virasoro algebra \cite{Faddeev:1993pe}, while the difference-operator
approach and its deformed Virasoro structures were developed in
refs.~\cite{FrenkelReshetikhin1996,FrenkelReshetikhinSemenov1997}.
In particular, the familiar two-factor Miura coefficient and its
nonlinear discrete Poisson bracket already belong to this mathematical
literature. We neither claim a new lattice Virasoro algebra nor regard
its continuum Virasoro limit, by itself, as a new result.

The missing ingredient is a dilaton realization compatible with that
\emph{same} finite-lattice Poisson structure. We work with an odd number
of periodic sites and a nonvanishing real Cartan zero mode. A single
multiplicative Cartan variable determines the ordered second-order
Miura operator and its lattice spin-two coefficient. We then select a
periodic scalar dilaton by solving an inhomogeneous first-order
difference equation associated with the Miura transfer problem. The
periodicity condition fixes the solution within the chosen nontrivial
monodromy sector. As a result, the dilaton becomes an explicit,
nonlocal function of the original free fields, with no additional
Poisson data prescribed on this Miura phase space.

This distinction has a direct algebraic consequence. Besides the
already known spin-two bracket, the common free-field Poisson tensor
induces a definite mixed bracket between the Miura current and the
composite dilaton. We derive this bracket at finite lattice spacing
and express its dependence on the periodic transport kernel. The
transfer-matrix description also relates the scalar construction to an
adjoint variable with a site-independent quadratic Casimir. These
statements concern the selected Miura realization; they do not promote
the composite scalar to an unrestricted, independent BF dilaton.

The central consistency test is the finite Fourier analysis and its
continuum limit. The lattice spacing is taken to zero at fixed
boundary circumference, with the Cartan zero mode and the Fourier
normalization retained throughout. The established spin-two sector
approaches the classical Virasoro structure. More importantly, the
induced mixed bracket must reproduce the continuous transformation of
the JT dilaton as a field of conformal weight minus one, in the
conventions of ref.~\cite{Ozer:2025bpb}. This correspondence is a
statement about the classical free-field and Ward structures, not a
quantization of the lattice theory.

The relation to gravitational charges must be stated separately. The
fundamental adjoint dilaton in BF gravity and the composite scalar in
the Miura realization belong to different starting descriptions.
Matching their transformation laws does not establish equality between
their symplectic structures, nor does it construct the integrable
boundary charges of the discrete BF theory. Such an identification
would require an explicit boundary variational principle and its
symplectic reduction. The present analysis is restricted to the
finite-periodic free-field realization and its controlled continuous
comparison; no higher-spin extension is assumed.

\Needspace{8\baselineskip}
The remainder of the paper is organized as follows.
Section~\ref{sec:bf-jt} recalls the continuous and discrete BF setting
and its conformal boundary data. Section~\ref{sec:discrete-miura}
summarizes the established discrete Miura--Virasoro construction and
its finite Fourier algebra. Section~\ref{sec:dilaton-free-field}
constructs the periodic dilaton from the same Cartan variables,
derives the induced mixed Poisson bracket and determines the associated
adjoint Casimir. Section~\ref{sec:fourier-continuum} develops the exact finite
Fourier realization of the mixed bracket and establishes its continuum
correspondence with the JT dilaton Ward transformation. Finally,
Section~\ref{sec:discussion-conclusions} discusses the scope of the
free-field result, its gravitational interpretation and the remaining
open issues.
%%%%%%%%%%%%%%%%%%%%%%%%%%%%%%%%%%%%%%%%%%%%%%%%%%%%%%%%%%%%%%%%%%%%%%%%%%%%%%%%%%%%%%%%%%
%%%%%%%%%%%%%%%%%%%%%%%%%%%%%%%%%%%%%%%%%%%%%%%%%%%%%%%%%%%%%%%%%%%%%%%%%%%%%%%%%%%%%%%%%%
%%%%%%%%%%      SECTION-II   %%%%%%%%%%%%%%%%%%%%%%%%%%%%%%%%%%%%%%%%%%%%%%%%%%%%%%%%%%%%%
%%%%%%%%%%%%%%%%%%%%%%%%%%%%%%%%%%%%%%%%%%%%%%%%%%%%%%%%%%%%%%%%%%%%%%%%%%%%%%%%%%%%%%%%%%
%%%%%%%%%%%%%%%%%%%%%%%%%%%%%%%%%%%%%%%%%%%%%%%%%%%%%%%%%%%%%%%%%%%%%%%%%%%%%%%%%%%%%%%%%%
%============================================================
% Section 2: gravitational setting, retained from the parent manuscript
% and shortened to the data used by the subsequent Miura construction.
%============================================================
\section{Jackiw--Teitelboim gravity: from continuous to discrete BF theory}
\label{sec:bf-jt}

As motivated in Section~\ref{sec:introduction}, the continuous
$\mathfrak{sl}(2,\mathbb R)$ BF description and its
holonomy-based discretization provide the gravitational reference
for the free-field construction developed below. We recall the bulk
equations, the conformal boundary transformation and the finite-link
adjoint transport, following refs.~\cite{Ozer:2025bpb,OzerFiliz:2026DiscreteJT}.
The distinction between a fundamental BF dilaton and its prospective
Miura free-field representative will be kept explicit. The continuous
sector is recalled in subsection~\ref{subsec:continuous-bf}, the
holonomy discretization in subsection~\ref{subsec:discrete-bf}, and
the conformal boundary data in subsection~\ref{subsec:boundary-miura}.
%%%%%%%%%%%%%%%%%%%%%%%%%%%%%%%%%%%%%%%%%%%%%%%%%%%%%%%%%%%%%%%%%%%%%%%%%%%%%%%%%%%%%%%%%%
%%%%%%%%%%      SECTION-II-1 %%%%%%%%%%%%%%%%%%%%%%%%%%%%%%%%%%%%%%%%%%%%%%%%%%%%%%%%%%%%%
%%%%%%%%%%%%%%%%%%%%%%%%%%%%%%%%%%%%%%%%%%%%%%%%%%%%%%%%%%%%%%%%%%%%%%%%%%%%%%%%%%%%%%%%%%
\subsection{Continuous \texorpdfstring{$\mathfrak{sl}(2,\mathbb R)$}{sl(2,R)} BF gravity}
\label{subsec:continuous-bf}

On an oriented two-dimensional manifold $\mathcal M$ with boundary,
the basic fields are a connection $A$ and an adjoint-valued scalar $X$,
both taking values in $\mathfrak{sl}(2,\mathbb R)$. In the trace convention
adopted here, the JT BF action is
\begin{equation}
 I_{\mathrm{BF}}[X,A]
 =\frac{k}{4\pi}\int_{\mathcal M}\mathrm{tr}(XF)+I_{bdy},
 \qquad F=dA+A\wedge A.
 \label{eq:bf-action}
\end{equation}
The boundary term $I_{bdy}$ depends on the chosen variational
principle. The bulk equations obtained from \eqref{eq:bf-action} are
\begin{equation}
 F=0,\qquad D_A X=dX+[A,X]=0.
 \label{eq:bf-eom}
\end{equation}
They are covariant under the infinitesimal gauge transformations
\begin{equation}
 \delta_\Lambda A=d\Lambda+[A,\Lambda],\qquad
 \delta_\Lambda X=[X,\Lambda],
 \quad \Lambda\in\mathfrak{sl}(2,\mathbb R).
 \label{eq:bf-gauge}
\end{equation}
The overall normalization of canonical boundary charges is not fixed
by \eqref{eq:bf-eom}; it depends on the boundary term and admissible
variations in \eqref{eq:bf-action}.

We employ the usual radial gauge,
\begin{equation}
 A=b^{-1}ab+b^{-1}db,\qquad X=b^{-1}xb,\qquad
 b=e^{\rho L_0},\qquad a=a_\tau(\tau)d\tau,
 \label{eq:bf-radial-gauge}
\end{equation}
where $\tau\sim\tau+2\pi$. In the radial gauge \eqref{eq:bf-radial-gauge}, the second equation in \eqref{eq:bf-eom} implies
$\partial_\tau x+[a_\tau,x]=0$. In particular, the choice
$\Lambda=x$ in \eqref{eq:bf-gauge} stabilizes the connection,
\begin{equation}
 \delta_{\Lambda=x}a_\tau
 =\partial_\tau x+[a_\tau,x]=0.
 \label{eq:bf-stabilizer}
\end{equation}

For the generators $[L_i,L_j]=(i-j)L_{i+j}$, we choose
\begin{equation}
 L_{+1}=\begin{pmatrix}0&0\\1&0\end{pmatrix},\quad
 L_0=\frac12\begin{pmatrix}1&0\\0&-1\end{pmatrix},\quad
 L_{-1}=\begin{pmatrix}0&-1\\0&0\end{pmatrix}.
 \label{eq:bf-sl2-basis}
\end{equation}
In the basis \eqref{eq:bf-sl2-basis}, the conformal Drinfeld--Sokolov (DS) connection has a single
state-dependent current $\mathcal L(\tau)$,
\begin{equation}
 a_\tau^{\mathrm{DS}}
 =L_{+1}+\alpha\mathcal L(\tau)L_{-1}
 =\begin{pmatrix}0&u(\tau)\\1&0\end{pmatrix},
 \qquad u=-\alpha\mathcal L.
 \label{eq:bf-cont-ds}
\end{equation}
Preserving the matrix form in \eqref{eq:bf-cont-ds} fixes the residual
parameter in terms of a free function $\epsilon(\tau)$:
\begin{equation}
 \Lambda_\epsilon
 =\begin{pmatrix}
   -\frac12\epsilon'&u\epsilon-\frac12\epsilon''\\
    \epsilon&\frac12\epsilon'
  \end{pmatrix}.
 \label{eq:bf-residual-gauge}
\end{equation}
Substitution into \eqref{eq:bf-gauge} gives the classical spin-two
transformation
\begin{equation}
 \delta_\epsilon u=\epsilon u'+2\epsilon'u-\frac12\epsilon'''.
 \label{eq:bf-virasoro-ward}
\end{equation}
For a covariantly constant adjoint dilaton, the lower-left entry
$\mathcal X$ determines its diagonal entries through
$x_{11}=-\mathcal X'/2$ and $x_{22}=\mathcal X'/2$.
The commutator in \eqref{eq:bf-gauge}, with the residual parameter
\eqref{eq:bf-residual-gauge}, therefore implies
\begin{equation}
 \delta_\epsilon\mathcal X
 =\epsilon\mathcal X'-\epsilon'\mathcal X.
 \label{eq:bf-dilaton-ward}
\end{equation}
Together with the spin-two law \eqref{eq:bf-virasoro-ward},
the transformation \eqref{eq:bf-dilaton-ward} characterizes a
conformal weight $-1$ and is the mixed-sector continuum test to be
used later. For a covariantly constant dilaton, the identification of
$x$ with the stabilizing parameter of \eqref{eq:bf-residual-gauge}
additionally gives
\begin{equation}
 \mathcal X'''-4u\mathcal X'-2u'\mathcal X=0.
 \label{eq:bf-dilaton-stabilizer}
\end{equation}
The stabilizer equation \eqref{eq:bf-dilaton-stabilizer} should not be
confused with the transformation law \eqref{eq:bf-dilaton-ward}:
the former fixes an allowed background, while the latter describes
how the dilaton transforms along the DS boundary phase space.
%%%%%%%%%%%%%%%%%%%%%%%%%%%%%%%%%%%%%%%%%%%%%%%%%%%%%%%%%%%%%%%%%%%%%%%%%%%%%%%%%%%%%%%%%%
%%%%%%%%%%      SECTION-II-2 %%%%%%%%%%%%%%%%%%%%%%%%%%%%%%%%%%%%%%%%%%%%%%%%%%%%%%%%%%%%%
%%%%%%%%%%%%%%%%%%%%%%%%%%%%%%%%%%%%%%%%%%%%%%%%%%%%%%%%%%%%%%%%%%%%%%%%%%%%%%%%%%%%%%%%%%
\subsection{Discrete BF action, holonomies and dilaton fields}
\label{subsec:discrete-bf}

Let $\mathcal K$ be an oriented cell complex with links $\ell$,
plaquettes $f$ and basepoints $b(f)$. The discrete connection is
represented by group-valued links $U_\ell\in SL(2,\mathbb R)$, with
$U_{\ell^{-1}}=U_\ell^{-1}$. For a link $\ell:s(\ell)\to t(\ell)$,
our endpoint convention reads
\begin{equation}
 U_\ell\longmapsto g_{s(\ell)}^{-1}U_\ell g_{t(\ell)},
 \qquad g_v\in SL(2,\mathbb R).
 \label{eq:bf-link-gauge}
\end{equation}
An ordered plaquette product
$W_f={\prod_{\ell\in\partial f}}U_\ell$
transforms by conjugation at $b(f)$. Near the group identity,
a holonomy-based discrete BF action is
\begin{equation}
 I^{\mathrm{disc}}_{\mathrm{BF}}
 =\frac{k}{4\pi}\sum_f\mathrm{tr}(X_f\log W_f)
 +I_{bdy,\Delta},
 \label{eq:bf-discrete-action}
\end{equation}
where the adjoint multipliers $X_f$ are based at $b(f)$. The
logarithm in \eqref{eq:bf-discrete-action} is understood on a local
branch. Variation with respect to $X_f$ gives local flatness,
\begin{equation}
 W_f=\mathbf 1.
 \label{eq:bf-plaquette-flatness}
\end{equation}
The local flatness condition \eqref{eq:bf-plaquette-flatness} does not allow arbitrary nontrivial holonomy around a
contractible loop on an unpunctured disk. Nontrivial global monodromy
requires a compatible topology, puncture or defect sector.

Link variation of \eqref{eq:bf-discrete-action} relates transported
\emph{face} multipliers. A site-based dilaton equation requires a
consistent identification with the appropriate dual transport data.
Under that identification its gauge-covariant form is
\begin{equation}
 X_{t(\ell)}=U_\ell^{-1}X_{s(\ell)}U_\ell.
 \label{eq:bf-adjoint-transport}
\end{equation}
The adjoint action in \eqref{eq:bf-adjoint-transport},
compatible with the link gauge law \eqref{eq:bf-link-gauge}, preserves
$\frac12\mathrm{tr}(X^2)$, the quadratic
$\mathfrak{sl}(2,\mathbb R)$ Casimir. These properties are
reviewed here only to identify the gravitational transport and
monodromy conditions relevant to our scalar free-field candidate.
%%%%%%%%%%%%%%%%%%%%%%%%%%%%%%%%%%%%%%%%%%%%%%%%%%%%%%%%%%%%%%%%%%%%%%%%%%%%%%%%%%%%%%%%%%
%%%%%%%%%%      SECTION-II-3 %%%%%%%%%%%%%%%%%%%%%%%%%%%%%%%%%%%%%%%%%%%%%%%%%%%%%%%%%%%%%
%%%%%%%%%%%%%%%%%%%%%%%%%%%%%%%%%%%%%%%%%%%%%%%%%%%%%%%%%%%%%%%%%%%%%%%%%%%%%%%%%%%%%%%%%%
\subsection{Conformal boundary data and the Miura--DS sector}
\label{subsec:boundary-miura}

We use a periodic boundary lattice with $L$ sites, spacing
$h=2\pi/L$, and the forward difference
\begin{equation}
 \tau_n=nh,\qquad
 (\nabla f)_n=\frac{f_{n+1}-f_n}{h},\qquad
 n\in\mathbb Z/L\mathbb Z.
 \label{eq:bf-boundary-lattice}
\end{equation}
The periodicity convention in \eqref{eq:bf-boundary-lattice} is the
same as that used for the Miura fields in
Section~\ref{sec:discrete-miura}. The present gravitational
kinematics does not require odd $L$; that restriction will enter
through the invertibility of the Miura Poisson kernel.

A boundary link $U_n$ has, in a local logarithm branch, the form
$U_n=\exp(h a_n)$, with $a_n\in\mathfrak{sl}(2,\mathbb R)$.
Its exact finite-link gauge variation and the corresponding adjoint
dilaton transport are
\begin{equation}
 \delta_\Lambda U_n=-\Lambda_nU_n+U_n\Lambda_{n+1},
 \qquad x_{n+1}=U_n^{-1}x_nU_n.
 \label{eq:bf-boundary-link-laws}
\end{equation}
Thus a covariantly constant adjoint dilaton determines a stabilizing
gauge parameter at finite spacing: setting $\Lambda_n=x_n$ in
\eqref{eq:bf-boundary-link-laws} gives $\delta_{\Lambda=x}U_n=0$, the finite-link counterpart
of the continuous stabilizer condition \eqref{eq:bf-stabilizer}.

The conformal boundary ansatz is the logarithmic DS restriction
\begin{equation}
 a_n^{\mathrm{DS}}
 =L_{+1}+\alpha\mathcal L_n L_{-1},\qquad
 U_n^{\mathrm{DS}}=\exp(h a_n^{\mathrm{DS}}).
 \label{eq:bf-discrete-ds}
\end{equation}
Only the spin-two field $\mathcal L_n$ occurs in
\eqref{eq:bf-discrete-ds}. This is a boundary-link ansatz, not a
proof that the resulting canonical BF charge bracket equals a
particular Miura Poisson bracket. Likewise, the $SL(2,\mathbb R)$
companion transfer matrix introduced below cannot be identified
with $U_n^{\mathrm{DS}}$ at finite spacing without a specific change
of basis and orientation convention. The next section constructs the
algebraic Miura side separately; the correspondence will be tested
through its Virasoro--dilaton transformation laws.
%%%%%%%%%%%%%%%%%%%%%%%%%%%%%%%%%%%%%%%%%%%%%%%%%%%%%%%%%%%%%%%%%%%%%%%%%%%%%%%%%%%%%%%%%%
%%%%%%%%%%%%%%%%%%%%%%%%%%%%%%%%%%%%%%%%%%%%%%%%%%%%%%%%%%%%%%%%%%%%%%%%%%%%%%%%%%%%%%%%%%
%%%%%%%%%%      SECTION-III  %%%%%%%%%%%%%%%%%%%%%%%%%%%%%%%%%%%%%%%%%%%%%%%%%%%%%%%%%%%%%
%%%%%%%%%%%%%%%%%%%%%%%%%%%%%%%%%%%%%%%%%%%%%%%%%%%%%%%%%%%%%%%%%%%%%%%%%%%%%%%%%%%%%%%%%%
%%%%%%%%%%%%%%%%%%%%%%%%%%%%%%%%%%%%%%%%%%%%%%%%%%%%%%%%%%%%%%%%%%%%%%%%%%%%%%%%%%%%%%%%%%
\section{Discrete Miura construction and Virasoro Poisson algebra}\label{sec:discrete-miura}

The gravitational variables of Section~\ref{sec:bf-jt} provide the
comparison framework; the two-factor Miura map and its nonlinear
lattice Virasoro bracket are established results of difference
Drinfeld--Sokolov reduction
\cite{Faddeev:1993pe,FrenkelReshetikhinSemenov1997}. We recall them in a
finite-periodic convention that will also be used for the composite
dilaton in the following dilaton analysis. In particular, the Poisson
structure is specified on the Cartan variables before any composite
field is introduced. No identification with the canonical BF boundary
bracket is required at this stage.
The ordered operator and its companion system are fixed in
subsection~\ref{subsec:cartan-miura}, the free-field bracket in
subsection~\ref{subsec:cartan-poisson}, and the exact finite
Fourier representation in subsection~\ref{subsec:miura-fourier}.
%%%%%%%%%%%%%%%%%%%%%%%%%%%%%%%%%%%%%%%%%%%%%%%%%%%%%%%%%%%%%%%%%%%%%%%%%%%%%%%%%%%%%%%%%%
%%%%%%%%%%      SECTION-III-1 %%%%%%%%%%%%%%%%%%%%%%%%%%%%%%%%%%%%%%%%%%%%%%%%%%%%%%%%%%%%
%%%%%%%%%%%%%%%%%%%%%%%%%%%%%%%%%%%%%%%%%%%%%%%%%%%%%%%%%%%%%%%%%%%%%%%%%%%%%%%%%%%%%%%%%%
\subsection{Periodic Cartan variables and the ordered Miura operator}\label{subsec:cartan-miura}

Let the boundary circle be represented by an odd number $L\geq3$ of
sites, with site labels $n\in\mathbb Z/L\mathbb Z$, spacing
$h=2\pi/L$, and $\tau_n=nh$. We use the forward shift $S$, defined by
$(Sf)_n=f_{n+1}$. For a site-dependent multiplication operator it
obeys $Sf_n=f_{n+1}S$, which fixes the ordering in the following
construction. Introduce a real periodic Cartan potential $\varphi_n$
and its constant mode $\nu$ through
\begin{equation}
 j_n=\nu+\frac{\varphi_{n+1}-\varphi_n}{h},\qquad
 y_n=h j_n,\qquad \lambda_n=e^{y_n}>0,
 \qquad \varphi_{n+L}=\varphi_n .
 \label{eq:cartan-vars}
\end{equation}
The total logarithmic Cartan mode is
\begin{equation}
 P=\sum_{n=0}^{L-1}y_n=2\pi\nu,
 \qquad \prod_{n=0}^{L-1}\lambda_n=e^P.
 \label{eq:cartan-zero-mode}
\end{equation}
The total mode in \eqref{eq:cartan-zero-mode} will be a Casimir
of the free-field bracket introduced below. The composite dilaton
considered subsequently requires $P\neq0$, whereas this Poisson
structure is defined for arbitrary $P$. The separation of the zero
mode and the nonzero Fourier modes in \eqref{eq:cartan-vars} is
essential for this distinction.

For the $\mathfrak{sl}(2,\mathbb R)$ Cartan sector we choose the ordered
second-order Miura operator
\begin{equation}
 \mathcal M_2=(S-\lambda_n)(S-\lambda_n^{-1})
   =S^2-t_n S+1,\qquad
 t_n=\lambda_n+\lambda_{n+1}^{-1}.
 \label{eq:miura-operator}
\end{equation}
The shifted inverse in the current defined by
\eqref{eq:miura-operator} follows from the operator ordering;
replacing it by $\lambda_n^{-1}$ would define a different difference
operator. Thus there is one nonconstant scalar coefficient, $t_n$,
which we use as the lattice spin-two current. The associated difference
equation and its companion transfer are
\begin{equation}
\begin{gathered}
 \psi_{n+2}-t_n\psi_{n+1}+\psi_n=0,\qquad
 v_{n+1}=C_n v_n,\qquad
 v_n=\begin{pmatrix}\psi_{n+1}\\ \psi_n\end{pmatrix},\\[3pt]
 C_n=\begin{pmatrix}t_n&-1\\1&0\end{pmatrix}\in SL(2,\mathbb R).
\end{gathered}
 \label{eq:miura-companion}
\end{equation}
The companion matrix in \eqref{eq:miura-companion} has unit determinant
and provides the transfer problem from which scalar solutions and
monodromy data will be constructed; it is not, on its own, an identification
of $C_n$ with a gravitational BF link.

For smooth $j(\tau)$, the continuum expansion at fixed $\tau_n$ is
\begin{equation}
 t_n=e^{h j_n}+e^{-h j_{n+1}}
   =2+h^2\bigl(j^2-j'\bigr)_{\tau_n}+O(h^3),
 \qquad u=j^2-j'.
 \label{eq:miura-continuum-potential}
\end{equation}
The expansion \eqref{eq:miura-continuum-potential} shows why the
background value $t_n=2$ must be subtracted before extracting $u$.
Accordingly, $h^{-2}\mathcal M_2$ approaches the differential operator
$(\partial_\tau-j)(\partial_\tau+j)=\partial_\tau^2-u$
on smooth test functions. This is the conventional continuous Miura
potential. The Poisson and Fourier limits require a separate scaling
analysis and will be treated together with the dilaton continuum limit.
%%%%%%%%%%%%%%%%%%%%%%%%%%%%%%%%%%%%%%%%%%%%%%%%%%%%%%%%%%%%%%%%%%%%%%%%%%%%%%%%%%%%%%%%%%
%%%%%%%%%%      SECTION-III-2 %%%%%%%%%%%%%%%%%%%%%%%%%%%%%%%%%%%%%%%%%%%%%%%%%%%%%%%%%%%%
%%%%%%%%%%%%%%%%%%%%%%%%%%%%%%%%%%%%%%%%%%%%%%%%%%%%%%%%%%%%%%%%%%%%%%%%%%%%%%%%%%%%%%%%%%
\subsection{The finite-periodic free-field Poisson bracket}\label{subsec:cartan-poisson}

We equip the logarithmic Cartan fields with the established periodic
Miura bracket
\begin{equation}
 \{y_n,y_m\}_{\mathrm M}=\phi_{n-m},\qquad
 \{\lambda_n,\lambda_m\}_{\mathrm M}
   =\phi_{n-m}\lambda_n\lambda_m,
 \label{eq:cartan-poisson}
\end{equation}
The kernel entering \eqref{eq:cartan-poisson} is defined
modulo the odd integer $L$ by
\begin{equation}
 \phi_0=0,\qquad \phi_d=(-1)^d\quad(1\leq d<L),
 \qquad \phi_{d+L}=\phi_d.
 \label{eq:alternating-kernel}
\end{equation}
The oddness assumption in \eqref{eq:alternating-kernel}
ensures $\phi_{-d}=-\phi_d$ and
$\sum_{m=0}^{L-1}\phi_{n-m}=0$. The constant Poisson tensor in
\eqref{eq:cartan-poisson} is antisymmetric and obeys the Jacobi
identity; in particular,
\begin{equation}
 \{P,y_m\}_{\mathrm M}=0.
 \label{eq:cartan-casimir}
\end{equation}
Equation~\eqref{eq:cartan-casimir}, together with the definition
\eqref{eq:cartan-zero-mode}, shows that the zero mode is a Casimir
of this chosen free-field bracket, rather than an additional independent dynamical current.

Applying the Leibniz rule to the two terms in $t_n$ gives
\begin{align}
 \{t_n,t_m\}_{\mathrm M}
 & =\phi_{n-m}\lambda_n\lambda_m
 -\phi_{n-m-1}\frac{\lambda_n}{\lambda_{m+1}}
 -\phi_{n+1-m}\frac{\lambda_m}{\lambda_{n+1}}
 +\phi_{n-m}\frac{1}{\lambda_{n+1}\lambda_{m+1}}.
 \label{eq:miura-bracket-expanded}
\end{align}
To simplify \eqref{eq:miura-bracket-expanded}, one uses the
following exact cyclic identities of the alternating kernel:
\begin{equation}
 \phi_{d-1}+\phi_d=\delta^{(L)}_{d,0}-\delta^{(L)}_{d,1},
 \qquad
 \phi_{d+1}+\phi_d=-\delta^{(L)}_{d,0}+\delta^{(L)}_{d,-1},
 \label{eq:kernel-localization}
\end{equation}
where $\delta^{(L)}$ denotes the periodic Kronecker delta. Subtracting
$\phi_{n-m}t_nt_m$ from \eqref{eq:miura-bracket-expanded} and using
\eqref{eq:kernel-localization}, the remaining terms collapse to
nearest-neighbor contributions. The closed result is
\begin{equation}
 \displaystyle
 \{t_n,t_m\}_{\mathrm M}=
 \phi_{n-m}t_nt_m
 +\delta^{(L)}_{n,m+1}-\delta^{(L)}_{n+1,m}.
 \label{eq:discrete-virasoro}
\end{equation}
Equation~\eqref{eq:discrete-virasoro} is the classical lattice
Virasoro bracket of
refs.~\cite{Faddeev:1993pe,FrenkelReshetikhinSemenov1997} in our
normalization. Its Jacobi identity follows directly from
\eqref{eq:cartan-poisson}, since the $t_n$ are functions of the
underlying Cartan fields. This is a known input, not a claimed new
Virasoro algebra.
%%%%%%%%%%%%%%%%%%%%%%%%%%%%%%%%%%%%%%%%%%%%%%%%%%%%%%%%%%%%%%%%%%%%%%%%%%%%%%%%%%%%%%%%%%
%%%%%%%%%%      SECTION-III-3 %%%%%%%%%%%%%%%%%%%%%%%%%%%%%%%%%%%%%%%%%%%%%%%%%%%%%%%%%%%%
%%%%%%%%%%%%%%%%%%%%%%%%%%%%%%%%%%%%%%%%%%%%%%%%%%%%%%%%%%%%%%%%%%%%%%%%%%%%%%%%%%%%%%%%%%
\subsection{Exact finite Fourier representation}\label{subsec:miura-fourier}

For the exact Fourier representation of
\eqref{eq:discrete-virasoro}, we fix the forward and inverse
conventions as follows:
\begin{equation}
 \zeta=e^{2\pi i/L},\qquad
 \widehat t_r=\frac{1}{L}\sum_{n=0}^{L-1}\zeta^{-rn}t_n,
 \qquad
 t_n=\sum_{r\in\mathbb Z/L\mathbb Z}\zeta^{rn}\widehat t_r.
 \label{eq:t-fourier}
\end{equation}
With the Fourier conventions of \eqref{eq:t-fourier}, the
alternating kernel in \eqref{eq:alternating-kernel} takes the form
\begin{equation}
 \phi_d=\frac1L\sum_{k=1}^{L-1}f_k\zeta^{kd},
 \qquad
 f_k=\frac{1-\zeta^{-k}}{1+\zeta^{-k}}
     =i\tan\!\left(\frac{\pi k}{L}\right).
 \label{eq:kernel-fourier}
\end{equation}
The denominator is nonzero because $L$ is odd; the constant mode is
absent. Substituting \eqref{eq:kernel-fourier} into the double Fourier
transform of \eqref{eq:discrete-virasoro} yields the exact nonlinear
mode bracket
\begin{equation}
 \displaystyle
 \{\widehat t_r,\widehat t_s\}_{\mathrm M}
 =\frac1L\sum_{k=1}^{L-1}
   c_k\widehat t_{r+k}\widehat t_{s-k}
 +\frac{\zeta^{-r}-\zeta^{-s}}{L}\,
    \delta^{(L)}_{r+s,0},\qquad
 c_k=\frac{1-\zeta^k}{1+\zeta^k}
     =-i\tan\!\left(\frac{\pi k}{L}\right).
 \label{eq:finite-fourier-virasoro}
\end{equation}
All mode indices in \eqref{eq:finite-fourier-virasoro} are
understood modulo $L$. In particular,
\eqref{eq:finite-fourier-virasoro} is an identity at finite spacing,
not a replacement of the discrete bracket by its continuum limit.

The same Fourier kernel defines the convolution operator
\begin{equation}
 (K_h f)_n=\sum_{m=0}^{L-1}\phi_{n-m}f_m,
 \label{eq:kernel-operator}
\end{equation}
which, by \eqref{eq:kernel-operator}, acts diagonally on
lattice Fourier waves:
\begin{equation}
 K_h\zeta^{\ell n}
 =i\tan\!\left(\frac{\pi\ell}{L}\right)\zeta^{\ell n}.
 \label{eq:kernel-symbol}
\end{equation}
For fixed Fourier mode $\ell$, the eigenvalue in
\eqref{eq:kernel-symbol} admits the low-momentum expansion as
$h\to0$,
\begin{equation}
 K_h=\frac h2\partial_\tau
       -\frac{h^3}{24}\partial_\tau^3+O(h^5),
 \label{eq:kernel-weak-expansion}
\end{equation}
understood on smooth functions rather than as a pointwise limit of
$\phi_d$. The expression \eqref{eq:kernel-weak-expansion}
is not an exact finite-lattice differential identity. Its role in the continuum Virasoro and
mixed dilaton limits will be examined after the composite dilaton
has been constructed.
%%%%%%%%%%%%%%%%%%%%%%%%%%%%%%%%%%%%%%%%%%%%%%%%%%%%%%%%%%%%%%%%%%%%%%%%%%%%%%%%%%%%%%%%%%
%%%%%%%%%%%%%%%%%%%%%%%%%%%%%%%%%%%%%%%%%%%%%%%%%%%%%%%%%%%%%%%%%%%%%%%%%%%%%%%%%%%%%%%%%%
%%%%%%%%%%      SECTION-IV   %%%%%%%%%%%%%%%%%%%%%%%%%%%%%%%%%%%%%%%%%%%%%%%%%%%%%%%%%%%%%
%%%%%%%%%%%%%%%%%%%%%%%%%%%%%%%%%%%%%%%%%%%%%%%%%%%%%%%%%%%%%%%%%%%%%%%%%%%%%%%%%%%%%%%%%%
%%%%%%%%%%%%%%%%%%%%%%%%%%%%%%%%%%%%%%%%%%%%%%%%%%%%%%%%%%%%%%%%%%%%%%%%%%%%%%%%%%%%%%%%%%
\section{Free-field realization of the dilaton}
\label{sec:dilaton-free-field}

The lattice current $t_n$ and its Poisson algebra were constructed in
Section~\ref{sec:discrete-miura} from the logarithmic Cartan variables
$y_n$.  We now use \emph{the same variables} to build a periodic dilaton
and to derive its mixed bracket with $t_n$.  The companion transfer
\eqref{eq:miura-companion} gives the required difference problem; no
additional Poisson tensor is prescribed.  Throughout this section $L$
is odd, $h=2\pi/L$, $\beta\in\mathbb R$ is fixed, and the Cartan
zero mode in \eqref{eq:cartan-zero-mode} belongs to the nonresonant
sector $P\ne0$.  First we construct a fundamental solution matrix in
subsection~\ref{subsec:dilaton-fundamental}; next we determine its
periodic adjoint representative in
subsection~\ref{subsec:dilaton-periodic}; finally we calculate the
induced mixed Poisson bracket in
subsection~\ref{subsec:dilaton-mixed}.
%%%%%%%%%%%%%%%%%%%%%%%%%%%%%%%%%%%%%%%%%%%%%%%%%%%%%%%%%%%%%%%%%%%%%%%%%%%%%%%%%%%%%%%%%%
%%%%%%%%%%      SECTION-IV-1  %%%%%%%%%%%%%%%%%%%%%%%%%%%%%%%%%%%%%%%%%%%%%%%%%%%%%%%%%%%%
%%%%%%%%%%%%%%%%%%%%%%%%%%%%%%%%%%%%%%%%%%%%%%%%%%%%%%%%%%%%%%%%%%%%%%%%%%%%%%%%%%%%%%%%%%
\subsection{Two Miura solutions and their monodromy}
\label{subsec:dilaton-fundamental}

The factorization \eqref{eq:miura-operator} immediately supplies one
nonvanishing solution of the second-order scalar equation.  Fixing its
normalization at site zero, we set
\begin{equation}
 \psi_0=1,\qquad
 \psi_{n+1}=\lambda_n^{-1}\psi_n,\qquad
 \psi_n=\exp\!\left(-\sum_{a=0}^{n-1}y_a\right).
 \label{eq:dilaton-psi}
\end{equation}
Indeed, the second first-order factor in $\mathcal M_2$ annihilates
$\psi_n$, so \eqref{eq:dilaton-psi} solves the companion problem
\eqref{eq:miura-companion}.  All expressions are extended to the
covering integer lattice using the periodic coefficients $\lambda_n$.

A second, independent solution is obtained by discrete reduction of
order.  Let $r_0=0$ and define
\begin{equation}
 r_{n+1}-r_n=\frac{h}{\psi_n\psi_{n+1}},\qquad
 \chi_n=\psi_n r_n,\qquad
 r_n=h\sum_{a=0}^{n-1}\frac{1}{\psi_a\psi_{a+1}}.
 \label{eq:dilaton-second-solution}
\end{equation}
The ratio in \eqref{eq:dilaton-second-solution} is fixed so that
$\psi_{n+1}\chi_n-\psi_n\chi_{n+1}=-h$.  Since the companion transfer has
unit determinant, this Wronskian is conserved.  The corresponding
fundamental matrix thus satisfies
\begin{equation}
 F_n=\begin{pmatrix}
       \psi_{n+1}&\chi_{n+1}\\
       \psi_n&\chi_n
      \end{pmatrix},\qquad
 F_{n+1}=C_nF_n,\qquad \det F_n=-h.
 \label{eq:dilaton-fundamental-matrix}
\end{equation}
In particular, the two columns of $F_n$ in
\eqref{eq:dilaton-fundamental-matrix} are independent solutions of the
\emph{same} Miura difference equation, not newly introduced fields.

The zero-mode identity \eqref{eq:cartan-zero-mode} implies
\begin{equation}
 \psi_{n+L}=e^{-P}\psi_n,\qquad
 r_{n+L}=R+e^{2P}r_n,\qquad
 R=h\sum_{a=0}^{L-1}\frac{1}{\psi_a\psi_{a+1}}.
 \label{eq:dilaton-monodromy-solutions}
\end{equation}
Equivalently, the solution basis has the monodromy
\begin{equation}
 F_{n+L}=F_n\mathsf D,\qquad
 \mathsf D=\begin{pmatrix}
              e^{-P}&e^{-P}R\\0&e^P
            \end{pmatrix},\qquad \det\mathsf D=1.
 \label{eq:dilaton-solution-monodromy}
\end{equation}
For real $P\ne0$, the two eigenvalues of $\mathsf D$ in
\eqref{eq:dilaton-solution-monodromy} are distinct.  This is the
hyperbolic Cartan monodromy sector selected by the positive Miura
factors; it is not a statement about all admissible BF holonomies.
%%%%%%%%%%%%%%%%%%%%%%%%%%%%%%%%%%%%%%%%%%%%%%%%%%%%%%%%%%%%%%%%%%%%%%%%%%%%%%%%%%%%%%%%%%
%%%%%%%%%%      SECTION-IV-2  %%%%%%%%%%%%%%%%%%%%%%%%%%%%%%%%%%%%%%%%%%%%%%%%%%%%%%%%%%%%
%%%%%%%%%%%%%%%%%%%%%%%%%%%%%%%%%%%%%%%%%%%%%%%%%%%%%%%%%%%%%%%%%%%%%%%%%%%%%%%%%%%%%%%%%%
\subsection{Periodic composite dilaton and adjoint Casimir}
\label{subsec:dilaton-periodic}

The solution monodromy \eqref{eq:dilaton-monodromy-solutions} singles out
a periodic scalar combination.  Define
\begin{equation}
 A=\frac{R}{e^{2P}-1},\qquad
\mathcal X_n=\beta\psi_n^2(r_n+A).
 \label{eq:dilaton-composite}
\end{equation}
Because $P\ne0$, the coefficient $A$ in
\eqref{eq:dilaton-composite} is finite.  Substituting
\eqref{eq:dilaton-monodromy-solutions} shows directly that
$\mathcal X_{n+L}=\mathcal X_n$.  More locally, the first-order
identity for $\psi_n$ and the increment of $r_n$ give the exact
dilaton equation
\begin{equation}
 \mathcal X_{n+1}
  =\lambda_n^{-2}\mathcal X_n+\beta h\lambda_n^{-1},
  \qquad \mathcal X_{n+L}=\mathcal X_n.
 \label{eq:dilaton-affine-transport}
\end{equation}
Its homogeneous multiplier over one period is $e^{-2P}\ne1$.
Consequently, \eqref{eq:dilaton-affine-transport} admits exactly one
periodic solution for the prescribed $\{y_n\}$ and $\beta$.
Iterating the equation backward around the circle yields an explicit
formula containing only the original Miura variables:
\begin{equation}
 \displaystyle
 \mathcal X_n=\frac{\beta h}{1-e^{-2P}}
 \sum_{\ell=0}^{L-1}
 \left(\prod_{j=1}^{\ell}\lambda_{n-j}^{-2}\right)
 \lambda_{n-1-\ell}^{-1}.
 \label{eq:dilaton-periodic-sum}
\end{equation}
All site indices in \eqref{eq:dilaton-periodic-sum} are taken modulo
$L$, and an empty product equals one.  The explicit expression also
proves that $\mathcal X_n$ depends on no free variable beyond those
already present in \eqref{eq:cartan-poisson}.  Its nonlocality at
finite spacing reflects the periodic transport condition, not an
additional dynamical degree of freedom.

The scalar construction also admits a precise adjoint realization.
Consider the constant traceless matrix
\begin{equation}
 H_\beta=\begin{pmatrix}
      -\beta/2&\beta A\\0&\beta/2
    \end{pmatrix},\qquad
 x_n^{\mathrm M}=F_nH_\beta F_n^{-1},\qquad
 [\mathsf D,H_\beta]=0.
 \label{eq:dilaton-adjoint-representative}
\end{equation}
The commutator in \eqref{eq:dilaton-adjoint-representative} vanishes
precisely because of the definition of $A$ in
\eqref{eq:dilaton-composite}.  Together with
\eqref{eq:dilaton-fundamental-matrix} and
\eqref{eq:dilaton-solution-monodromy}, it gives
\begin{equation}
 x_{n+1}^{\mathrm M}=C_nx_n^{\mathrm M}C_n^{-1},\qquad
 x_{n+L}^{\mathrm M}=x_n^{\mathrm M},\qquad
 \mathcal X_n=h\,(x_n^{\mathrm M})_{21}.
 \label{eq:dilaton-adjoint-transport}
\end{equation}
The factor $h$ in the last equality of
\eqref{eq:dilaton-adjoint-transport} follows from the Wronskian
normalization $\det F_n=-h$; it is needed to recover exactly the scalar
normalization in \eqref{eq:dilaton-composite}.  Conjugation also gives
the site-independent quadratic invariant
\begin{equation}
 \frac12\operatorname{tr}\!\bigl[(x_n^{\mathrm M})^2\bigr]
 =\frac12\operatorname{tr}(H_\beta^2)
 =\frac{\beta^2}{4}.
 \label{eq:dilaton-casimir}
\end{equation}
Thus the periodic scalar in \eqref{eq:dilaton-periodic-sum} is
compatible with the nonzero adjoint Casimir
\eqref{eq:dilaton-casimir} when $\beta\ne0$.  The
matrix $x_n^{\mathrm M}$ is an adjoint representative of the Miura
companion transport.  With the endpoint convention of
\eqref{eq:bf-boundary-link-laws}, the associated candidate link would
have orientation $C_n^{-1}$.  We do not assume that it already equals
the gravitational DS link \eqref{eq:bf-discrete-ds}, or that its
induced Poisson structure equals the BF boundary-charge bracket.
%%%%%%%%%%%%%%%%%%%%%%%%%%%%%%%%%%%%%%%%%%%%%%%%%%%%%%%%%%%%%%%%%%%%%%%%%%%%%%%%%%%%%%%%%%
%%%%%%%%%%      SECTION-IV-3  %%%%%%%%%%%%%%%%%%%%%%%%%%%%%%%%%%%%%%%%%%%%%%%%%%%%%%%%%%%%
%%%%%%%%%%%%%%%%%%%%%%%%%%%%%%%%%%%%%%%%%%%%%%%%%%%%%%%%%%%%%%%%%%%%%%%%%%%%%%%%%%%%%%%%%%
\subsection{The induced mixed Miura--dilaton Poisson bracket}
\label{subsec:dilaton-mixed}

We now compute the mixed bracket rather than prescribing its form.
For compactness, introduce the site variables
\begin{equation}
 a_n=\lambda_n^{-2},\qquad b_n=\lambda_n^{-1},\qquad
 D_n=2a_n\mathcal X_n+\beta h b_n
     =\mathcal X_{n+1}+a_n\mathcal X_n.
 \label{eq:dilaton-mixed-definitions}
\end{equation}
Using the definitions \eqref{eq:dilaton-mixed-definitions},
differentiating the exact transport equation
\eqref{eq:dilaton-affine-transport} with respect to a logarithmic
Cartan field $y_b$ yields the inhomogeneous linear recurrence
\begin{equation}
 \frac{\partial\mathcal X_{n+1}}{\partial y_b}
 -a_n\frac{\partial\mathcal X_n}{\partial y_b}
 =-\delta^{(L)}_{n,b}\,D_n.
 \label{eq:dilaton-variational-recurrence}
\end{equation}
Because $\mathcal X_n$ is the periodic function
\eqref{eq:dilaton-periodic-sum}, its partial derivatives are periodic
as well.  We solve \eqref{eq:dilaton-variational-recurrence} using
cyclic transport.  Set
\begin{equation}
 \begin{aligned}
 d(m,b)&=(m-b-1)\!\!\pmod L,\qquad 0\le d(m,b)<L,\\
 G_{mb}&=\frac{1}{1-e^{-2P}}
            \prod_{j=1}^{d(m,b)}a_{b+j},\\
 \frac{\partial\mathcal X_m}{\partial y_b}&=-G_{mb}D_b.
 \end{aligned}
 \label{eq:dilaton-green-kernel}
\end{equation}
Here the product in \eqref{eq:dilaton-green-kernel} follows the
forward cyclic path starting at site $b+1$ and stops just before
$m$; in particular $G_{b+1,b}=(1-e^{-2P})^{-1}$.  The denominator is
nonzero in the sector $P\ne0$.  Direct substitution into
\eqref{eq:dilaton-variational-recurrence} verifies the Green-kernel
identity, including the step across the cyclic source.

The current in \eqref{eq:miura-operator} has the elementary derivatives
\begin{equation}
 \frac{\partial t_k}{\partial y_a}
 =\lambda_k\delta^{(L)}_{k,a}
 -\lambda_{k+1}^{-1}\delta^{(L)}_{k+1,a}.
 \label{eq:dilaton-current-derivative}
\end{equation}
Applying the free-field Poisson tensor
\eqref{eq:cartan-poisson} to the derivatives in
\eqref{eq:dilaton-green-kernel} and
\eqref{eq:dilaton-current-derivative}, we obtain the central
finite-lattice result of this section:
\begin{equation}
 \begin{aligned}
 \{t_k,\mathcal X_m\}_{\mathrm M}
 =-\sum_{b=0}^{L-1}\Bigl[
 &\lambda_k\phi_{k-b}
 -\lambda_{k+1}^{-1}\phi_{k+1-b}\Bigr]\,
 G_{mb}D_b.
 \end{aligned}
 \label{eq:dilaton-mixed-poisson}
\end{equation}
Equation~\eqref{eq:dilaton-mixed-poisson} is an \emph{induced} bracket:
every quantity on the right-hand side is a function of the original
Cartan fields, whose brackets were fixed in
\eqref{eq:cartan-poisson}.  The alternating kernel $\phi$ carries
the Miura Poisson structure, while $G$ carries the globally periodic
dilaton transport.  Their combination gives a finite-lattice mixed
algebra without adding a second set of canonical variables.

The algebraic consistency of \eqref{eq:dilaton-mixed-poisson}
follows directly from its construction.  The field $t_n$ and the
composite $\mathcal X_m$ are differentiable functions of the same
finite collection of $y_a$, and the latter carry a constant Poisson
tensor.  Hence the mixed Jacobi identity, for example
\begin{equation}
 \{t_k,\{t_\ell,\mathcal X_m\}_{\mathrm M}\}_{\mathrm M}
 +\{t_\ell,\{\mathcal X_m,t_k\}_{\mathrm M}\}_{\mathrm M}
 +\{\mathcal X_m,\{t_k,t_\ell\}_{\mathrm M}\}_{\mathrm M}=0,
 \label{eq:dilaton-mixed-jacobi}
\end{equation}
is inherited from the free-field bracket
\eqref{eq:cartan-poisson}; the identity
\eqref{eq:dilaton-mixed-jacobi} is not a new closure constraint to be
imposed on \eqref{eq:dilaton-mixed-poisson}.  The centrality of $P$
in \eqref{eq:cartan-casimir} also guarantees that the dependence of
the periodic Green kernel on $e^{-2P}$ introduces no extra Poisson
data.  The exact Fourier transform and continuous Ward comparison of
the mixed bracket are derived in Section~\ref{sec:fourier-continuum}.
%%%%%%%%%%%%%%%%%%%%%%%%%%%%%%%%%%%%%%%%%%%%%%%%%%%%%%%%%%%%%%%%%%%%%%%%%%%%%%%%%%%%%%%%%%
%%%%%%%%%%%%%%%%%%%%%%%%%%%%%%%%%%%%%%%%%%%%%%%%%%%%%%%%%%%%%%%%%%%%%%%%%%%%%%%%%%%%%%%%%%
%%%%%%%%%%      SECTION-V    %%%%%%%%%%%%%%%%%%%%%%%%%%%%%%%%%%%%%%%%%%%%%%%%%%%%%%%%%%%%%
%%%%%%%%%%%%%%%%%%%%%%%%%%%%%%%%%%%%%%%%%%%%%%%%%%%%%%%%%%%%%%%%%%%%%%%%%%%%%%%%%%%%%%%%%%
%%%%%%%%%%%%%%%%%%%%%%%%%%%%%%%%%%%%%%%%%%%%%%%%%%%%%%%%%%%%%%%%%%%%%%%%%%%%%%%%%%%%%%%%%%
\section{Finite Fourier algebra and the continuum JT limit}
\label{sec:fourier-continuum}

The finite-lattice Virasoro bracket was fixed in
Section~\ref{sec:discrete-miura}, while
Section~\ref{sec:dilaton-free-field} constructed a periodic scalar
$\mathcal X_n$ from precisely the same logarithmic Cartan variables.
The mixed Poisson bracket \eqref{eq:dilaton-mixed-poisson} is therefore
already determined. We now transform it without replacing the finite
lattice by a continuum algebra and then take a controlled limit. The
exact mode realization is given in
subsection~\ref{subsec:mixed-fourier}; the Ward transformations follow
from the free-field Hamiltonian flow in
subsection~\ref{subsec:continuum-ward}; their mode brackets and
comparison with the continuous JT data are discussed in
subsection~\ref{subsec:jt-fourier-comparison}.
Throughout, $L$ is odd, $h=2\pi/L$, the field $j$ is smooth and
periodic, and its fixed Cartan zero mode satisfies
$P=\int_0^{2\pi} j(\tau)\,d\tau\ne0$. The parameter $\beta$ is held
fixed. Continuum Fourier statements refer to fixed mode numbers as
$L\to\infty$; no termwise limit of the alternating site kernel is used.
%%%%%%%%%%%%%%%%%%%%%%%%%%%%%%%%%%%%%%%%%%%%%%%%%%%%%%%%%%%%%%%%%%%%%%%%%%%%%%%%%%%%%%%%%%
%%%%%%%%%%      SECTION- V-1  %%%%%%%%%%%%%%%%%%%%%%%%%%%%%%%%%%%%%%%%%%%%%%%%%%%%%%%%%%%%
%%%%%%%%%%%%%%%%%%%%%%%%%%%%%%%%%%%%%%%%%%%%%%%%%%%%%%%%%%%%%%%%%%%%%%%%%%%%%%%%%%%%%%%%%%
\subsection{Exact finite Fourier realization of the mixed bracket}
\label{subsec:mixed-fourier}

In addition to the spin-two modes of \eqref{eq:t-fourier}, define the
finite Fourier coefficients of the composite dilaton and the mixed
bracket by
\begin{equation}
 \widehat{\mathcal X}_s
 =\frac1L\sum_{m=0}^{L-1}\zeta^{-sm}\mathcal X_m,
 \qquad
 \mathcal Q_{rs}^{(L)}
 =\{\widehat t_r,\widehat{\mathcal X}_s\}_{\mathrm M}.
 \label{eq:mixed-fourier-definitions}
\end{equation}
All Fourier labels in \eqref{eq:mixed-fourier-definitions} are
understood modulo $L$. Direct substitution of the exact site bracket
\eqref{eq:dilaton-mixed-poisson} gives
\begin{equation}
 \begin{aligned}
 \mathcal Q_{rs}^{(L)}=-\frac1{L^2}
 \sum_{k,m,b=0}^{L-1}\zeta^{-rk-sm}
 \bigl(\lambda_k\phi_{k-b}
      -\lambda_{k+1}^{-1}\phi_{k+1-b}\bigr)
 G_{mb}D_b .
 \end{aligned}
 \label{eq:mixed-fourier-exact-sum}
\end{equation}
Equation~\eqref{eq:mixed-fourier-exact-sum} holds for every odd
$L\geq3$ and every real periodic $\{y_n\}$ with $P\ne0$; in
particular, it does not assume a slowly varying background.
Although the site kernel $\phi$ is nonlocal, its Fourier symbol is
known exactly from \eqref{eq:kernel-fourier}.

To exhibit that symbol explicitly, introduce the partial Fourier
transform of the periodic Green kernel and the corresponding composite
coefficients,
\begin{equation}
 \begin{gathered}
 \widehat G_s(b)=\frac1L\sum_{m=0}^{L-1}
      \zeta^{-sm}G_{mb},\\[2pt]
 B_b^{(s)}=D_b\widehat G_s(b),\qquad
 \widehat B_p^{(s)}=\frac1L\sum_{b=0}^{L-1}
      \zeta^{-pb}B_b^{(s)}.
 \end{gathered}
 \label{eq:mixed-fourier-green-symbols}
\end{equation}
For any periodic field $f_n$ we write
$\widehat f_q=L^{-1}\sum_n\zeta^{-qn}f_n$, including
$\widehat{\lambda^{-1}}_q$ for the inverse Cartan field. The
convolution with $\phi$ in \eqref{eq:mixed-fourier-exact-sum} can
then be evaluated using the spectral coefficients
$f_p=i\tan(\pi p/L)$ from \eqref{eq:kernel-fourier}. One obtains
the equivalent exact expression
\begin{equation}
 \displaystyle
 \mathcal Q_{rs}^{(L)}
 =-\sum_{p=1}^{L-1}f_p
 \left(\widehat\lambda_{r-p}
       -\zeta^{r}\widehat{\lambda^{-1}}_{r-p}\right)
 \widehat B_p^{(s)},\qquad
 f_p=i\tan\!\left(\frac{\pi p}{L}\right).
 \label{eq:mixed-fourier-spectral}
\end{equation}
To check the phase in \eqref{eq:mixed-fourier-spectral}, the shifted
inverse contribution in \eqref{eq:mixed-fourier-exact-sum} is
relabelled by $k+1\mapsto k$, producing precisely $\zeta^r$.
The intermediate coefficients in
\eqref{eq:mixed-fourier-green-symbols} are fixed entirely by the
periodic dilaton construction. Equations~\eqref{eq:mixed-fourier-exact-sum}
and \eqref{eq:mixed-fourier-spectral} are two forms of the same finite
Fourier realization. The coefficients $\widehat B_p^{(s)}$ depend on
the original Cartan field through the periodic transport kernel;
there is no assertion that the finite lattice bracket becomes a
local polynomial expression solely in the two displayed mode families.
%%%%%%%%%%%%%%%%%%%%%%%%%%%%%%%%%%%%%%%%%%%%%%%%%%%%%%%%%%%%%%%%%%%%%%%%%%%%%%%%%%%%%%%%%%
%%%%%%%%%%      SECTION- V-2  %%%%%%%%%%%%%%%%%%%%%%%%%%%%%%%%%%%%%%%%%%%%%%%%%%%%%%%%%%%%
%%%%%%%%%%%%%%%%%%%%%%%%%%%%%%%%%%%%%%%%%%%%%%%%%%%%%%%%%%%%%%%%%%%%%%%%%%%%%%%%%%%%%%%%%%
\subsection{Continuum free-field flow and dilaton Ward identity}
\label{subsec:continuum-ward}

To keep the normalization explicit, extract the continuum spin-two
variable from the shifted Miura coefficient and introduce the
lattice-smeared Hamiltonian
\begin{equation}
 u_n^{(h)}=\frac{t_n-2}{h^2},\qquad
 H_h[\epsilon]=h\sum_{n=0}^{L-1}\epsilon_n u_n^{(h)}
 =\frac1h\sum_{n=0}^{L-1}\epsilon_n(t_n-2),
 \qquad \epsilon_n=\epsilon(\tau_n).
 \label{eq:continuum-hamiltonian}
\end{equation}
The subtraction of the constant background in
\eqref{eq:continuum-hamiltonian} does not change the induced
Hamiltonian flow. Using the free-field bracket
\eqref{eq:cartan-poisson}, the exact lattice variation of the Cartan
current is
\begin{equation}
 \displaystyle
 \delta_\epsilon j_m
 =\{j_m,H_h[\epsilon]\}_{\mathrm M}
 =\frac{1}{h^2}\,
 K_h\!\left[\epsilon_n\lambda_n
           -\epsilon_{n-1}\lambda_n^{-1}\right]_m.
 \label{eq:cartan-exact-flow}
\end{equation}
Here $K_h$ is the operator \eqref{eq:kernel-operator}; brackets in
\eqref{eq:cartan-exact-flow} indicate its argument, not an additional
Poisson operation. In particular, the total logarithmic mode remains
fixed by \eqref{eq:cartan-casimir}, so the flow stays in the same
$P\ne0$ sector.

For samples of smooth periodic functions, the factor in square
brackets in \eqref{eq:cartan-exact-flow} has the expansion
\begin{equation}
 \epsilon_n\lambda_n
 -\epsilon_{n-1}\lambda_n^{-1}
 =h\bigl(\epsilon'+2\epsilon j\bigr)_{\tau_n}
 +O(h^2).
 \label{eq:cartan-flow-input-expansion}
\end{equation}
Combining \eqref{eq:cartan-flow-input-expansion} with the
low-momentum operator expansion \eqref{eq:kernel-weak-expansion}
gives, for fixed smooth Fourier data,
\begin{equation}
 \lim_{h\to0}\delta_\epsilon j
 =(\epsilon j)'+\frac12\epsilon''.
 \label{eq:cartan-continuum-ward}
\end{equation}
The continuum Miura relation \eqref{eq:miura-continuum-potential},
$u=j^2-j'$, then implies directly
\begin{equation}
 \displaystyle
 \delta_\epsilon u
 =\epsilon u'+2\epsilon'u-\frac12\epsilon'''.
 \label{eq:virasoro-continuum-ward}
\end{equation}
Thus the central-derivative normalization of the classical spin-two
transformation follows from the finite free-field bracket; it has
not been inserted as an independent transformation rule.

The lattice dilaton has its own controlled continuum limit. Expanding
the exact inhomogeneous recurrence
\eqref{eq:dilaton-affine-transport} gives the periodic first-order
equation
\begin{equation}
 \displaystyle
 \mathcal X'+2j\mathcal X=\beta,
 \qquad \mathcal X(\tau+2\pi)=\mathcal X(\tau).
 \label{eq:dilaton-continuum-equation}
\end{equation}
Its homogeneous monodromy is $e^{-2P}$, so
\eqref{eq:dilaton-continuum-equation} has exactly one periodic
solution when $P\ne0$, just as at finite spacing. The solution
produced by \eqref{eq:dilaton-periodic-sum} converges to that
periodic solution under the smooth-profile assumptions stated above.

We can now derive, rather than postulate, the dilaton variation.
Keeping $\beta$ fixed and varying
\eqref{eq:dilaton-continuum-equation} yields
\begin{equation}
 (\partial_\tau+2j)\,\delta_\epsilon\mathcal X
 =-2\mathcal X\,\delta_\epsilon j.
 \label{eq:dilaton-linearized-equation}
\end{equation}
Insert the Cartan transformation
\eqref{eq:cartan-continuum-ward} into
\eqref{eq:dilaton-linearized-equation}. Differentiation of
\eqref{eq:dilaton-continuum-equation} shows that the candidate
$\epsilon\mathcal X'-\epsilon'\mathcal X$ satisfies exactly the
same inhomogeneous equation. Since the periodic homogeneous solution
is trivial for $P\ne0$, uniqueness establishes
\begin{equation}
 \displaystyle
 \delta_\epsilon\mathcal X
 =\epsilon\mathcal X'-\epsilon'\mathcal X.
 \label{eq:dilaton-continuum-ward}
\end{equation}
This derivation uses the same Hamiltonian flow as
\eqref{eq:virasoro-continuum-ward}; the scalar dilaton has conformal
weight $-1$ with no independent transformation law assigned to it.
%%%%%%%%%%%%%%%%%%%%%%%%%%%%%%%%%%%%%%%%%%%%%%%%%%%%%%%%%%%%%%%%%%%%%%%%%%%%%%%%%%%%%%%%%%
%%%%%%%%%%      SECTION- V-3  %%%%%%%%%%%%%%%%%%%%%%%%%%%%%%%%%%%%%%%%%%%%%%%%%%%%%%%%%%%%
%%%%%%%%%%%%%%%%%%%%%%%%%%%%%%%%%%%%%%%%%%%%%%%%%%%%%%%%%%%%%%%%%%%%%%%%%%%%%%%%%%%%%%%%%%
\subsection{Fourier Ward brackets and the continuous JT comparison}
\label{subsec:jt-fourier-comparison}

For fixed integers $r,s$ in the large-$L$ limit, define the
continuum Fourier coefficients of $u$ and $\mathcal X$ by
\begin{equation}
 \widehat u_r=\frac1{2\pi}\int_0^{2\pi}
      e^{-ir\tau}u(\tau)\,d\tau,
 \qquad
 \widehat{\mathcal X}_s=\frac1{2\pi}\int_0^{2\pi}
      e^{-is\tau}\mathcal X(\tau)\,d\tau.
 \label{eq:continuum-fourier-definitions}
\end{equation}
The lattice coefficients \eqref{eq:t-fourier} and
\eqref{eq:mixed-fourier-definitions} approach
\eqref{eq:continuum-fourier-definitions} through
$\widehat u_r^{(h)}=h^{-2}(\widehat t_r-2\delta^{(L)}_{r,0})$
and $\widehat{\mathcal X}_s^{(h)}=L^{-1}\sum_n\zeta^{-sn}\mathcal X_n$.
With the Hamiltonian normalization
\eqref{eq:continuum-hamiltonian}, the limiting generator reads
\begin{equation}
 H[\epsilon]=\int_0^{2\pi}\epsilon(\tau)u(\tau)\,d\tau
 =2\pi\sum_{r\in\mathbb Z}
      \widehat\epsilon_{-r}\widehat u_r.
 \label{eq:continuum-smeared-generator}
\end{equation}
With the generator \eqref{eq:continuum-smeared-generator},
Fourier transforming the two Ward transformations
\eqref{eq:virasoro-continuum-ward} and
\eqref{eq:dilaton-continuum-ward}, using
$\delta_\epsilon F=\{F,H[\epsilon]\}$, fixes the limiting
mode brackets:
\begin{equation}
 \begin{aligned}
 \{\widehat u_r,\widehat u_s\}_{\mathrm{cont}}
  &=\frac{i(r-s)}{2\pi}\widehat u_{r+s}
       +\frac{ir^3}{4\pi}\delta_{r+s,0},\\[2pt]
 \{\widehat u_r,\widehat{\mathcal X}_s\}_{\mathrm{cont}}
  &=-\frac{i(2r+s)}{2\pi}\widehat{\mathcal X}_{r+s}.
 \end{aligned}
 \label{eq:continuum-fourier-brackets}
\end{equation}
In particular, the second equality in
\eqref{eq:continuum-fourier-brackets} is the limit of the
\emph{exact} finite-lattice mixed realization
\eqref{eq:mixed-fourier-exact-sum}:
\begin{equation}
 \displaystyle
 \lim_{h\to0}\frac1{h^2}
 \{\widehat t_r,\widehat{\mathcal X}_s\}_{\mathrm M}
 =-\frac{i(2r+s)}{2\pi}\widehat{\mathcal X}_{r+s}.
 \label{eq:mixed-fourier-continuum-limit}
\end{equation}
The limit in \eqref{eq:mixed-fourier-continuum-limit} is taken on
smooth periodic Cartan profiles with fixed nonzero $P$, fixed $\beta$
and fixed Fourier labels, rather than on arbitrary high-frequency
lattice configurations. The spin-two equality in
\eqref{eq:continuum-fourier-brackets} is the familiar classical
Virasoro limit of the known difference algebra, whereas the second
one is the induced dilaton check of the present construction.

In position space, the continuum transformations
\eqref{eq:virasoro-continuum-ward} and
\eqref{eq:dilaton-continuum-ward} reproduce the classical
Virasoro and weight-minus-one dilaton Ward structures of the
continuous JT BF analysis \cite{Ozer:2025bpb}. To display both
singular structures in the conventional normalization of that
reference, let
\begin{equation}
\mathcal T=-\frac{k}{2}u,\qquad
 \{F,G\}_{k}=-\frac{2}{k}\{F,G\}_{\mathrm{cont}},
 \qquad k\ne0.
 \label{eq:ward-comparison-normalization}
\end{equation}
This is a rescaling of the classical continuum Miura Poisson
bracket for comparison purposes: by construction,
$\{F,\int\epsilon\mathcal T\}_{k}=\{F,\int\epsilon u\}_{\mathrm{cont}}$
for every Miura observable $F$. In particular, the two
transformations become
\begin{equation}
 \begin{aligned}
 \delta_\epsilon\mathcal T
   &=\epsilon \mathcal T'+2\epsilon'\mathcal T+\frac{k}{4}\epsilon''',\\
 \delta_\epsilon\mathcal X
   &=\epsilon\mathcal X'-\epsilon'\mathcal X.
 \end{aligned}
 \label{eq:ward-comparison-variations}
\end{equation}
In conventional classical Ward notation the corresponding
singular structures can therefore be displayed together as
\begin{subequations}\label{eq:classical-ope-comparison}
\begin{align}
\mathcal T(\tau_1)\mathcal T(\tau_2)&\sim
 \frac{3k}{2(\tau_1-\tau_2)^4}
 +\frac{2\mathcal T(\tau_2)}{(\tau_1-\tau_2)^2}
 +\frac{\partial_{\tau_2}\mathcal T(\tau_2)}
       {\tau_1-\tau_2},
 \label{eq:virasoro-classical-ope-comparison}\\
\mathcal T(\tau_1)\mathcal X(\tau_2)&\sim
 -\frac{\mathcal X(\tau_2)}{(\tau_1-\tau_2)^2}
 +\frac{\partial_{\tau_2}\mathcal X(\tau_2)}
       {\tau_1-\tau_2}.
 \label{eq:mixed-classical-ope-comparison}
\end{align}
\end{subequations}
The coefficient $3k/2$ in
\eqref{eq:virasoro-classical-ope-comparison} is the classical
Ward-normalization counterpart of the $k\epsilon'''/4$ term
in \eqref{eq:ward-comparison-variations}; it should not be
read as an independent derivation of the canonical BF central
charge from the Miura phase space. Equations
\eqref{eq:classical-ope-comparison} compare classical
transformation coefficients and do not assert quantum
operator-product identities. The gravitational statement and
the free-field result thus share the Virasoro--dilaton Ward
structure, while their canonical symplectic interpretations
remain distinct. In particular, equality between the Miura
bracket and the BF boundary-charge bracket does not follow
solely from this continuum match. The exact finite-lattice
realization \eqref{eq:mixed-fourier-spectral}, the
monodromy-selected dilaton \eqref{eq:dilaton-composite}, and
the controlled limit \eqref{eq:mixed-fourier-continuum-limit}
together provide the claimed Virasoro--dilaton free-field
correspondence.
%%%%%%%%%%%%%%%%%%%%%%%%%%%%%%%%%%%%%%%%%%%%%%%%%%%%%%%%%%%%%%%%%%%%%%%%%%%%%%%%%%%%%%%%%%
%%%%%%%%%%%%%%%%%%%%%%%%%%%%%%%%%%%%%%%%%%%%%%%%%%%%%%%%%%%%%%%%%%%%%%%%%%%%%%%%%%%%%%%%%%
%%%%%%%%%%      SECTION-VI   %%%%%%%%%%%%%%%%%%%%%%%%%%%%%%%%%%%%%%%%%%%%%%%%%%%%%%%%%%%%%
%%%%%%%%%%%%%%%%%%%%%%%%%%%%%%%%%%%%%%%%%%%%%%%%%%%%%%%%%%%%%%%%%%%%%%%%%%%%%%%%%%%%%%%%%%
%%%%%%%%%%%%%%%%%%%%%%%%%%%%%%%%%%%%%%%%%%%%%%%%%%%%%%%%%%%%%%%%%%%%%%%%%%%%%%%%%%%%%%%%%%
%%%%%%%%%%%%%%%%%%%%%%%%%%%%%%%%%%%%%%%%%%%%%%%%%%%%%%%%%%%%%%%%%%%%%%%%%%%%%%%%%%%%%%%%%%
%%%%%%%%%%%%%%%%%%%%%%%%%%%%%%%%%%%%%%%%%%%%%%%%%%%%%%%%%%%%%%%%%%%%%%%%%%%%%%%%%%%%%%%%%%
%%%%%%%%%%      DISCUSSION & CONCLUSIONS  %%%%%%%%%%%%%%%%%%%%%%%%%%%%%%%%%%%%%%%%%%%%%%%%
%%%%%%%%%%%%%%%%%%%%%%%%%%%%%%%%%%%%%%%%%%%%%%%%%%%%%%%%%%%%%%%%%%%%%%%%%%%%%%%%%%%%%%%%%%
%%%%%%%%%%%%%%%%%%%%%%%%%%%%%%%%%%%%%%%%%%%%%%%%%%%%%%%%%%%%%%%%%%%%%%%%%%%%%%%%%%%%%%%%%%
\section{Discussion and conclusions}
\label{sec:discussion-conclusions}

The purpose of this work was not to reconstruct the established lattice
Virasoro algebra, but to determine whether its Cartan free fields also
support a dilaton with the transformation law required by the
continuous JT boundary analysis. We distinguish the finite-lattice
result in subsection~\ref{subsec:discussion-results} from its
gravitational interpretation and outstanding restrictions in
subsection~\ref{subsec:discussion-scope}.
%%%%%%%%%%%%%%%%%%%%%%%%%%%%%%%%%%%%%%%%%%%%%%%%%%%%%%%%%%%%%%%%%%%%%%%%%%%%%%%%%%%%%%%%%%
%%%%%%%%%%      SECTION-VI-1 %%%%%%%%%%%%%%%%%%%%%%%%%%%%%%%%%%%%%%%%%%%%%%%%%%%%%%%%%%%%%
%%%%%%%%%%%%%%%%%%%%%%%%%%%%%%%%%%%%%%%%%%%%%%%%%%%%%%%%%%%%%%%%%%%%%%%%%%%%%%%%%%%%%%%%%%
\subsection{The finite-lattice Virasoro--dilaton realization}
\label{subsec:discussion-results}

The starting point is the known ordered Miura map
\eqref{eq:miura-operator}, together with the finite-periodic Cartan
Poisson structure \eqref{eq:cartan-poisson}. Their induced spin-two
bracket \eqref{eq:discrete-virasoro} is a standard result of the
difference Drinfeld--Sokolov construction
\cite{Faddeev:1993pe,FrenkelReshetikhinSemenov1997}. It provides the
algebraic framework here rather than an independent novelty claim.

The additional result is the composite dilaton
\eqref{eq:dilaton-periodic-sum}. It is fixed by the first-order
transport equation \eqref{eq:dilaton-affine-transport} and its periodic
monodromy condition, and depends only on the Cartan variables already
present in the Miura map. Its companion-matrix representative obeys
the adjoint transport law \eqref{eq:dilaton-adjoint-transport} and has
the constant quadratic invariant \eqref{eq:dilaton-casimir}. These
properties apply to the selected nonresonant monodromy sector; they do
not supply a second independent lattice current.

The mixed bracket \eqref{eq:dilaton-mixed-poisson} is the central
algebraic output. It follows from the same free-field Poisson tensor,
with the alternating lattice kernel and the periodic Green function
appearing together. The exact Fourier expressions
\eqref{eq:mixed-fourier-exact-sum} and
\eqref{eq:mixed-fourier-spectral} retain this transport dependence at
finite spacing. In particular, they do not imply a local polynomial
mode algebra involving only the displayed spin-two and dilaton
coefficients. For smooth configurations at fixed nonzero Cartan zero
mode, the limiting mixed bracket
\eqref{eq:mixed-fourier-continuum-limit} reproduces the weight-minus-one
Ward transformation \eqref{eq:dilaton-continuum-ward}. This agreement
provides the specific continuous JT check of the construction.
%%%%%%%%%%%%%%%%%%%%%%%%%%%%%%%%%%%%%%%%%%%%%%%%%%%%%%%%%%%%%%%%%%%%%%%%%%%%%%%%%%%%%%%%%%
%%%%%%%%%%      SECTION-VI-2 %%%%%%%%%%%%%%%%%%%%%%%%%%%%%%%%%%%%%%%%%%%%%%%%%%%%%%%%%%%%%
%%%%%%%%%%%%%%%%%%%%%%%%%%%%%%%%%%%%%%%%%%%%%%%%%%%%%%%%%%%%%%%%%%%%%%%%%%%%%%%%%%%%%%%%%%
\subsection{Gravitational scope and outlook}
\label{subsec:discussion-scope}

The BF description recalled in Section~\ref{sec:bf-jt} starts from an
independent adjoint dilaton and group-valued links. By contrast,
Section~\ref{sec:dilaton-free-field} builds a particular adjoint
representative and scalar component from a restricted Miura transfer
problem. Their matching covariant-transport and Ward properties do not
establish a canonical equivalence of phase spaces. In particular, no
boundary symplectic reduction, integrable BF boundary charge or
coupling-dependent identification of Poisson normalizations has been
constructed here. Such a derivation is necessary before interpreting
the induced Miura bracket as the gravitational asymptotic symmetry
algebra itself. The topology and admissible holonomy sectors of the BF
geometry must also be specified independently of the scalar transfer
problem.

The result has a defined domain. The finite Fourier construction uses
an odd number of sites; its alternating kernel has an additional
resonant mode for even lattice sizes. The periodic composite dilaton
is fixed in the real sector $P\ne0$, while the resonant sector
requires a separate solvability analysis. The continuum comparison
holds for smooth periodic profiles and fixed Fourier indices, not for
an unrestricted ultraviolet limit. The Ward correspondence is
classical and does not amount to a quantization of the dilaton or a
derivation of a quantum operator algebra.

Within these assumptions, the main outcome is a common discrete
free-field realization of the spin-two current and its dilaton
partner, with an explicitly induced mixed Poisson bracket and a
controlled JT Ward limit. The known Miura--Virasoro structure is thus
supplemented by a monodromy-selected dilaton in the same finite
Poisson space. Establishing the precise map to canonical BF boundary
charges, and extending the construction beyond the present periodic
sector, remain separate problems.
%%%%%%%%%%%%%%%%%%%%%%%%%%%%%%%%%%%%%%%%%%%%%%%%%%%%%%%%%%%%%%%%%%%%%%%%%%%%%%%%%%%%%%%%%%
%%%%%%%%%%%%%%%%%%%%%%%%%%%%%%%%%%%%%%%%%%%%%%%%%%%%%%%%%%%%%%%%%%%%%%%%%%%%%%%%%%%%%%%%%%
%%%%%%%%%%      BIBLIOGRAPHY  %%%%%%%%%%%%%%%%%%%%%%%%%%%%%%%%%%%%%%%%%%%%%%%%%%%%%%%%%%%%
%%%%%%%%%%%%%%%%%%%%%%%%%%%%%%%%%%%%%%%%%%%%%%%%%%%%%%%%%%%%%%%%%%%%%%%%%%%%%%%%%%%%%%%%%%
%%%%%%%%%%%%%%%%%%%%%%%%%%%%%%%%%%%%%%%%%%%%%%%%%%%%%%%%%%%%%%%%%%%%%%%%%%%%%%%%%%%%%%%%%%

\end{document}